\documentclass[aps,prl,reprint,groupedaddress]{revtex4-1}
\usepackage{graphicx}
\usepackage{physics}
\usepackage{dcolumn}
\usepackage{bm}
\usepackage{amsthm}
\newtheorem{post}{Postulate}
\newtheorem{ther}{Theorem}

\begin{document}


\title{Time Dilation as Quantum Tunneling Time}

\thanks{Work supported the URA., Inc., under contract DE-AC02-76CH03000 with the U.S. Dept. of Energy.}
\author{V. H.  Ranjbar}

\affiliation{Brookhaven National Lab, Upton NY 11973}


\date{\today}

\begin{abstract}
We conjecture that the
relative rate of time evolution depends on the amount of quantum
correlations in a system.  This is motivated by the experimental work \cite{PhysRevLett.119.023201} which showed that quantum
tunneling is not instantaneous. The non-zero tunneling
time may have other profound implications for the regulation of time
in an entangled system. It opens the possibility that other types of quantum
correlations may require non-zero rates of update. If this is true, it
provides a mechanism for regulating the relative rate of time
evolution.
\end{abstract}


\maketitle

\section{Introduction}
The formulation of General relativity leaves open the question of why
clocks tic relatively slower in the presence of higher
mass-energy. While it is understood as due to the deformation of
space-time by the stress-energy tensor, the microscopic  mechanism
or quantum mechanical basis for this is not understood. In this work we provide a possible explanation based on the number of
quantum correlated members of a given sytem. 

 Work by Van Raamsdonk \cite{VanRaamsdonk:2010pw} in consideration of
 the AdS/CFT correspondence has shown in the anti-de Sitter (AdS)
 toy universe, that space-time is a result of quantum
 entanglement on the conformal boundary. As a result it has been
 theorized that space-time is somehow built-up from quantum
 entanglement \cite{nature} and that this might resolve issues related
 to black holes and information \cite{Maldacena:2013xja} .  Prior to this Jacobson had demonstrated an approach to
 derive Einstein's equation of state from thermodynamic considerations
 \cite{PhysRevLett.75.1260}.  More recently, there have been attempts
 to derive General relativity using  entanglement entropy
 \cite{Lashkari:2013koa}, \cite{PhysRevLett.114.221601},  \cite{Verlinde:2016toy}.  

Related to this have been efforts to study
progress of time, computation and information storage as a product of quantum
mechanical processes. For example, there is the classical work by Jacob Bekenstein
\cite{Bekenstein:1973ur} showing that the theoretical
limit on information storage for a given space is governed by the
mass-energy of a black hole. Then there is the work by Norman
Margolus and Lev Levitin \cite{Margolus:1997ih} showing that a quantum system
with average energy E can evolve over a maximum of $ \frac{2 E}{\pi
  \hbar}$ states in a second.  This rate also happens to be the rate
of increase of space on the interior of a black hole as shown by Brown
\cite{Brown:2015lvg}. This work also relates the growth of this space to the
growth of computational complexity.  The fact that growth of space is regulated by
the growth of complexity, suggests that time should likewise be
regulated by some aspect of computational complexity.

That the progress of time could arise from quantum correlations was
recently tested \cite{Moreva:2013ska} showing that internal
correlation can behave as clocks while maintaining a static system as
viewed externally. This showed that the static universe provided by
the Wheeler-De Witt equations \cite{PhysRev.160.1113}  may still
permit the observation of time internally, perhaps in a manner similar
to the approach to a local
time suggested by \cite{Kitada:1994dq}.

Thus I am prompted to consider
entanglement complexity as a possible candidate for regulating the
rate of time evolution for a given mass-energy system. The recent
experimental evidence \cite{PhysRevLett.119.023201} that
quantum tunneling is not instantaneous opens the door to the
possibility that updates to quantum correlations may also not occur
instantaneously. If this is true then the relative complexity of quantum
correlations in a given system should slow the progress of time.

In its simplest understanding, the progress of time represents a
change of state for any given system. If a physical system has no
discernable change of state, time cannot be said to progress. This
change to a 'distinct' state can be clearly defined in terms of the
orthogonality
of the quantum state. So the evolution of a quantum state to an
orthogonal state represents the minimum process necessary to observe
an update in time.  Norman Margolus and Lev Levitin used this fact to
find the minimum time for a quantum state to evolve into an
orthogonal state to be $ \frac{h}{4 E} $.  This represents the
fastest time update rate possible, given non-interacting quantum state
functions. As the authors pointed out in their paper, it is analogous
to the maximum computation rate for a trivially parallel system. Here 
energy, if discretized could be thought of as representing the number
of processors. Incidentally if the orthogonality condition is removed
other quantum computational bounds can be derived \cite{DUGIC2002291}
which sharpen the distinction between ``classical'' and ``quantum'' information.

We argue in this paper that in reality physical systems don't behave in a
trivially parallel manner since they are entangled with the
surrounding environment.  For an entangled system any update of
state requires the update of all entangle partners to preserve
unitarity. So for example if the state of one particle of a simple
maximally entangle pair is collapsed via measurement, its partner must
also collapse. To avoid violations of unitarity it is has been assumed that
this collapse must occur instantaneously accross all entangle
pairs. However the results from tunneling experiments \cite{PhysRevLett.119.023201} might lead one to reconsider this
view. This work has shown that quantum tunneling effects do not occur
instantaneously. Tunneling times of 80-100 attosecs
were measured for their system. Tunneling can in some sense be
understood as the collapse of a superposition of two spatial location
for a particle. The wave function represents the probability that a
particle can exist in various locations. For a particle 
with a finite barrier interposing itself on the wave function, some of
those locations will be outside of the barrier and some inside.
 Thus it can be said to exist in a superposition of being behind the
 barrier and outside of it.
 The collapse of this superposition is what is measured when
 tunneling time is measured.
 Given this, one might expect that the collapse of a state function for entangled states also
wouldn't occur instantaneously. Generally this could imply that the
update to quantum mechanical state information requires a non-zero
time.
The question of non-zero collapse time for an entangled pair can and
should be settled by experiment as it was done for quantum tunneling time.
If this is true  then we have a mechanism which
could explain the microscopic relative behavior of time in a higher
mass-energy location.

In the case of two entangled pairs the only way to prevent violations of unitarity would be if an update to the physical state of
the of the second state were delayed by some finite amount to permit
the whole state to move into a net orthogonal state. Thus inorder to preserve unitarity of any given
system all updates to the physical state of a system must be
regulated by the number of entangled or quantum correlated partners. 

In this paper we re-derive the gravitational time-dilation
formula starting from the quantum tunneling time equation. In this
derivation we connect the quantum tunneling time to a measure of the rate
of propagation of quantum information through a bulk mass-energy system. 

\section{Entanglement and Tensor Networks}
The possibility that the collapse of entangled pairs doesn't occur
instantaneously but with a finite time, provides an avenue for
the a rate of time evolution to be determined by the
complexity of the operator.  This appears consistent with the view
that computational complexity provides for the growth of space behind
the black-hole horizon. This also provides a mechanism for the 
emergence of the force of gravity.

Another way of stating this is that the relative rate of time is
proportional to the number of parallel computational steps required to update
the state from one distinguishable state to another. In the case that
no entanglement exists, one would recover the  Margolus and Levitin
limit of  $ \frac{2 E}{\pi \hbar}$. But if the states are entangled
one with another then this rate is reduced by the fact that all
entangled chains will need to be updated before any state can evolve
into one orthogonal to it's current state. 

To quantify this better we propose our first postulate:
\begin{post}
 Information about the collapse of the quantum state function
  takes a non-zero time to propagate between entangled states.
\end{post}

\begin{figure}[h]
\center
\includegraphics*[width=75mm,height=40mm]{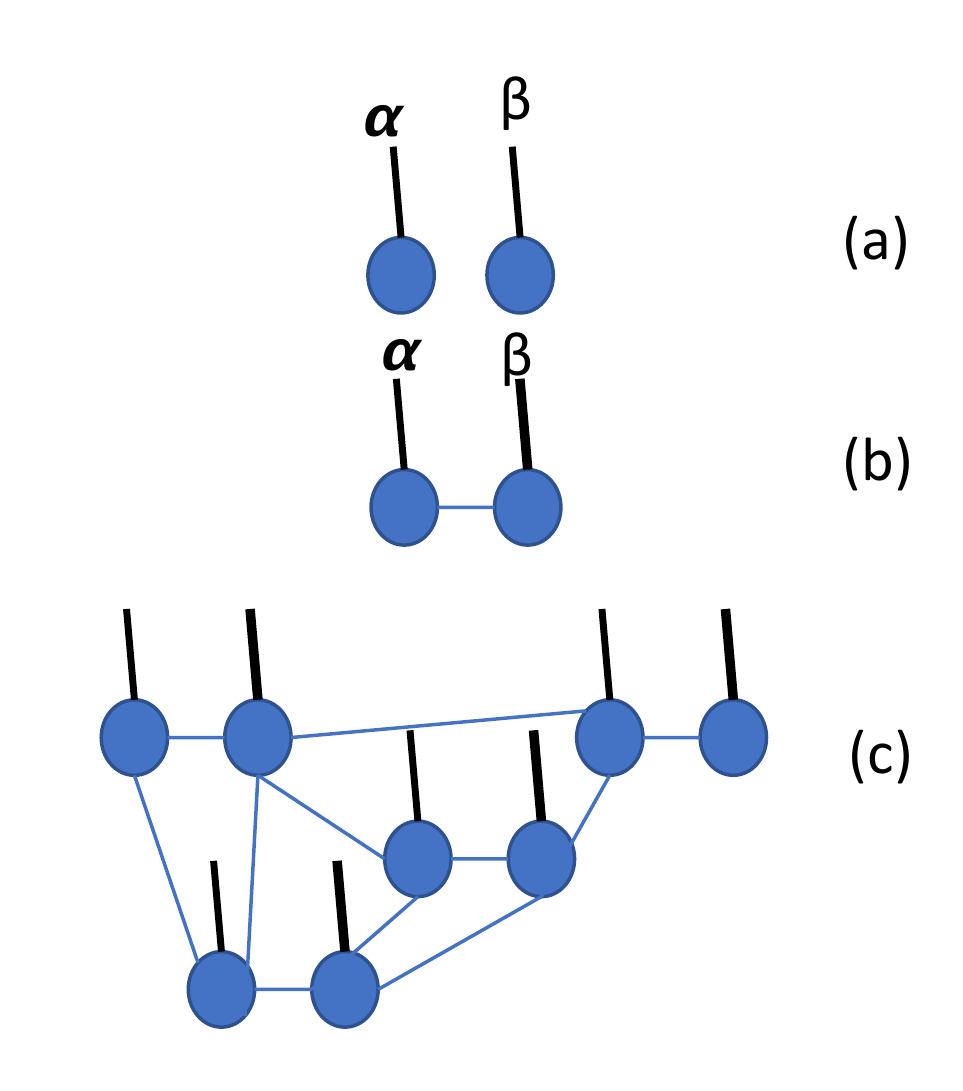}
\caption{Tensor Network Representing entanglement bonds. (a)
  represents unentangled state. (b) two entangled qubits, (c) several
  entangled qubits }
\label{fig1}
\end{figure}

To better understand how this would effect the evolution of time in an
entangled system we consider a tensor network representation of entanglement.
This approach is used to understand how space-time is built-up from
entanglement. Here a 
quantum many-body wave function is represented by nodes which are
connected to each other via entanglement. For example in
Fig.~\ref{fig1} we show in (a) a simple two-qubit state which is
unentangled (b) entangled and (c) a multi-qubit state with
entanglement. In this representation the blue nodes are the state
function $\ket{\psi} = a \ket{0} + b \ket{1}$ in this case a simple
one-qubit state as a linear combination of basis states. The vertical black lines
represent points at which you can calculate an amplitude
(i.e. $\bra{\alpha}\ket{\psi}$). The internal blue legs mediate
entanglement between the qubits.
Thus Fig.~\ref{fig1} (b) is
\begin{equation}
\bra{\alpha \beta}\ket{\psi} = \sum^{\chi}_{j=1} X^{\alpha}_j
Y^{\beta}_j ,
\end{equation}
here the range of the internal leg is $\chi$, the bond dimension.

In the case that the effects of entangled pairs require a
finite time to interact, then one could calculate the time it would
take for this information to propagate across a manybody wave function
like in Fig.~\ref{fig1} (c). This could be estimated using knowledge
of $\chi$ the bond dimension and the total number of bonds.

This leads to the first theorem following from our first postulate:

\begin{ther}
The maximum rate of update for a bulk entangled state is
  proportional to the total number and dimensions of the bonds as
  represented by it's tensor network.
\end{ther}

This theorem follows from the fact for the bulk state to evolve into a
distinct state different from its current state, it must evolve into
one which is orthogonal to it. If the state is not in a distinct
state from its current state, then time cannot be said to have
advanced for that system. For example if the initial state of  our many body entangled wave
function is represented by $\psi_0$ then for it to update to future
time $\psi_t$ they must be orthogonal to each other,
\begin{equation}
\bra{\psi_0}\ket{\psi_t} =0
\end{equation}
where 
\begin{equation}
\ket{\psi_t} =  \sum_{n} c_n e^{-i \frac{E_n t}{\hbar}} \ket{E_n}.
\end{equation}
 Before this process can be complete all information such as a
 collapse of one member via measurement must
have reached all nodes in the tensor network which are connected
otherwise unitarity will be violated.

  We introduce  a $\delta t_{min}$  which represents the minimum number of orthogonal
updates in an observing `non-entangled' system compared to
the entangled-system. By our conjecture, it is how many time tics in
the non-gravitational system necessary to observe one tic in the
gravitating system. A consequence of Theorem 1, is that this minimum
will grow with the size of the tensor network and number and size of
the entangled bonds.

\section{ Relationship between Mass-Energy and Quantum Correlations}

It has been proposed that number of entangled chains of particles is
proportional  to the mass-energy of a given system:
\begin{eqnarray}
N_{E} \propto E .
\label{eq1}
\end{eqnarray}
Here $N_{E}$ represents the number of entangled bonds. This is implied by \cite{PhysRevLett.114.221601} which relates
energy to the entanglement entropy.
It is also the conclusion of \cite{Leichenauer:2018obf} and a
consequence of the Quantum Null
Energy Condition which relates the stress energy tensor to Von Neuman entropy
via:
\begin{eqnarray}
S_{vv}^{\prime \prime} \le 2 \pi \langle T_{v v} \rangle . \label{eq2}
\end{eqnarray}

Qualitatively it has been shown \cite{Linden:2009gfy} that for a physical system the
introduction of a new member or particle causes a 
thermodynamic equilibration process which generates the formation of
quantum entanglement with the system.
Thus Eq.~\ref{eq1} is justified since each additional particle which becomes locally
entangled with a given system and will add another bond in the tensor
network for the whole system. It should be obvious that more
mass-energy leads to more potential particles from a field-theory
point of view and more bonds of entanglement will be created.

\begin{ther}
The total mass-energy of a given system is proportional to the amount
of entanglement as represented by the number and dimension of the
bonds in a tensor network representation
\end{ther}

\section{Tunneling time as time dilation?}

Calculating $\delta t_{min}$ by counting up all the bonds and their
dimensions in a given tensor network is challenging since we do not
yet know how to quantify how the time delay scales with each
additional bond and its dimension.

A clue to the dependence of the time delay on total mass-energy might be found in the tunneling time experimental and
theoretical work\cite{PhysRevLett.119.023201},
\cite{PhysRevB.27.6178}, \cite{Landsman:14},
\cite{PhysRevLett.53.115}.  Experimental work appears consistent with the
theoretical form given by the Feynman Path Integral formulation and
Larmor time approach. 
\begin{figure}[h]
\center
\includegraphics*[width=75mm,height=40mm]{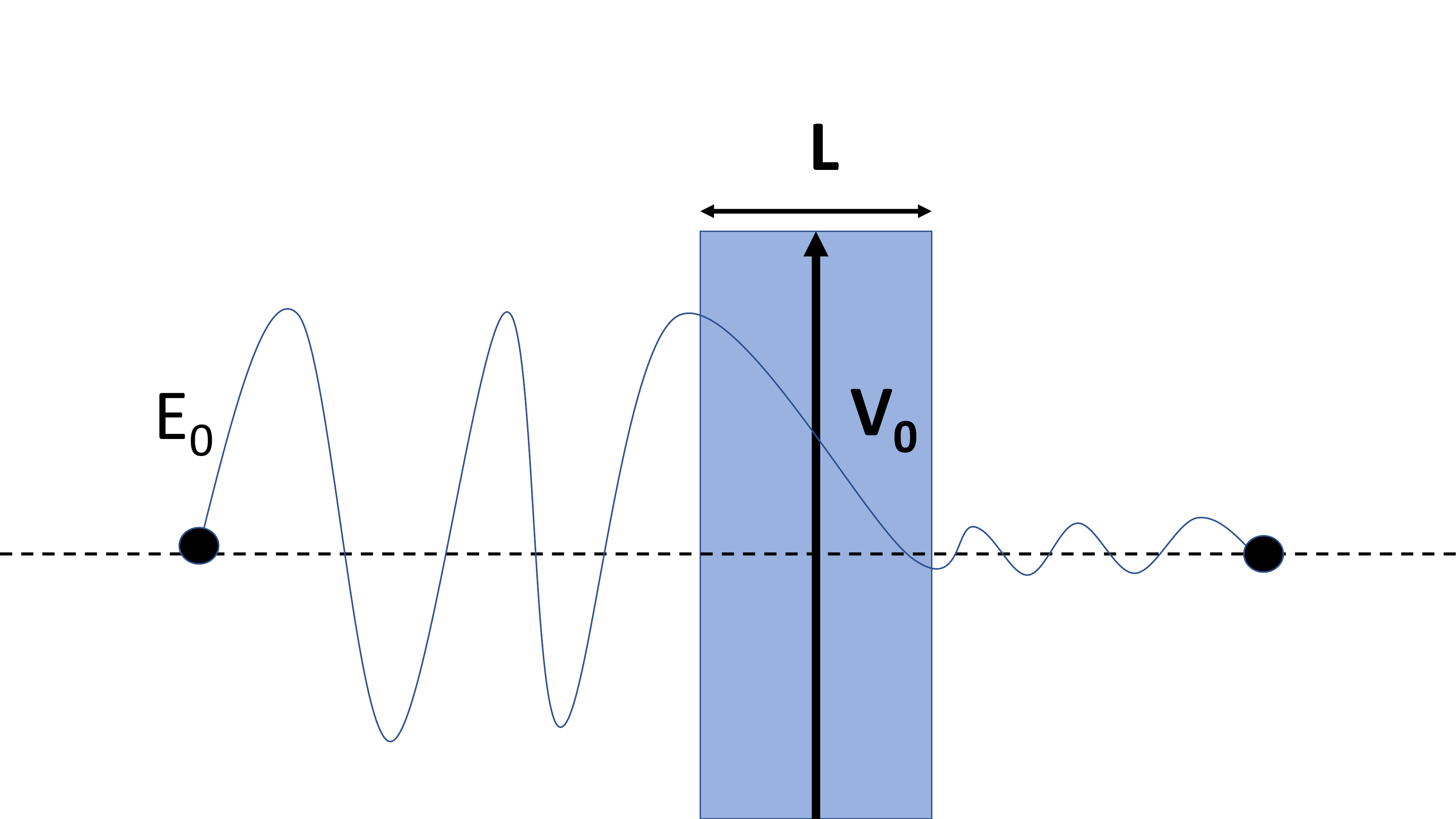}
\caption{Standard tunneling example through a potential barrier $V_0$
  of length $L$ and
incident particle of mass $m$ and energy $E_0$}
\label{fig2}
\end{figure}

Both give a functional form for the delay time
as:
\begin{eqnarray}
 t_T = \hbar \frac{\partial \ln |T(E_o)| }{ \partial E_0}. \label{eq4}
\end{eqnarray}
Here $t_T$ is the tunneling time and $T(E_0)$ is the transmission
probability for an incident particle of energy $E_0$. 
The functional form of $T(E)$ is given by the nature of the tunneling
barrier. The transmission coefficient for crossing a constant
potential barrier $V_0$, over a distance $L$ is given by  (see Fig.~\ref{fig2}) :
\begin{eqnarray}
T(E_0) = e^{-2 \sqrt{\frac{2m}{\hbar^2} ( V_0 - E_0)} L}.
\end{eqnarray}
Here $m$ is the rest mass of the tunneling particle.
Now applying Eq.~\ref{eq4}:
\begin{eqnarray}
t_T  & = & \frac{ \frac{ L m}{ \hbar} }{\sqrt{\frac{2 m}{\hbar^2} ( V_0 - E_0)}}
           \nonumber \\
& = & \frac{ 1}{\sqrt{\frac{1 }{2 L^2 m } ( V_0 - E_0)} }. \label{eq5}
\end{eqnarray}
One obtains the time spent under the potential barrier. Since the
particle is excluded from existing inside of the barrier, the
calculation of  tunneling time might be understood as a calculation
of the time it takes for quantum state information to propagate
through a bulk mass-energy system. This encapsulates the second
postulate:
\begin{post}
The transmission time for quantum
information through a bulk mass-energy system is proportional to  the
quantum tunneling time through the system. 
\end{post}

 It seems appropriate
to guess that information about the update to a particle's quantum
state follows a similar calculation.

Indeed in this form it looks suspiciously similar to the standard time
dilation form in the presence of mass-energy. One might imagine time
dilation as a result of the time delay for the propagation of quantum
state information across a total bulk mass-energy given by $E_T$. 
\begin{figure}[h]
\center
\includegraphics*[width=75mm,height=40mm]{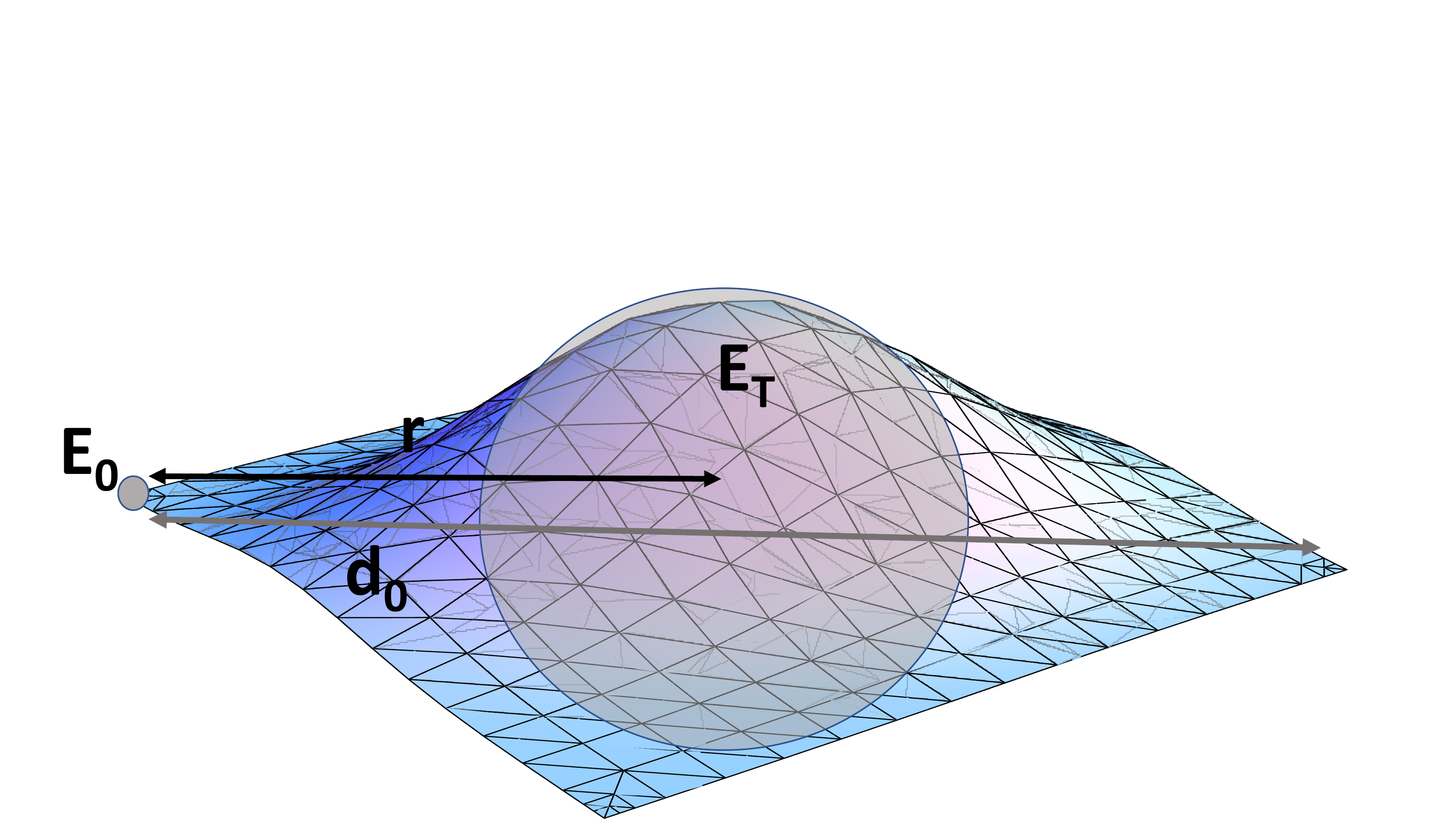}
\caption{Formulating time dilation as tunneling time. $E_T$ is the
  total mass-energy of the gravitating object (outlined in grey sphere), $r$ the distance to the
center of it's mass, $E_0$ the test particle's mass-energy and $d_o$
the distance tunneled (causal distance). The grid is the tensor
network or space-time with a bulge  representing more network bonds
caused by the concentration of mass-energy $E_T$. }
\label{fig3}
\end{figure}

if we assume the potential which is being tunneled
through is a linear function of the total mass-energy of the object
$V_0=a E_T=a M c^2$, (see Fig.~\ref{fig3})
 across a length $L=d_0$. Here $a$ is some yet to be determined
 porpotionality constant. Then by analogy to Eq.~\ref{eq5}, we define a
 transmission coefficient for the state collapse
$T_c$:
\begin{eqnarray}
T_c(E_0)= e^{-2 d_0 \sqrt{\frac{2 m_0}{\hbar^2} ( a M c^2 - E_0)} } .
\end{eqnarray}

Here $E_0$ and $m_0$ now represents the kinetic energy and rest mass of test particle approaching
an object of potential $aM c^2$ respectively. One could interpret $E_0$
and $m_0$  as the minimal kinetic energy and rest mass of the vacuum. 

$d_0$ for the cases of calculating
transmission coefficients is fixed for a fixed time scale, a sort of minimum
distance for a given time duration, beyond which quantum correlation updates are assumed to
be unnecessary to preserve unitarity. In
some sense the project of building up space-time from entanglement
means that 'distances' between things are given by the level of
entanglement of the space around them. So by definition two patches of
space at a distance $d$ from each other is more entangled than two
patches of space at a distance $2d$ from each other.  So
for example $d_0$ represents some average quantum correlation distance
of space, representing the natural fall from maximal entanglement at
larger distances. Thus beyond a certain average distance correlations
are so minimal that they don't contain information necessary for
updates to preserve unitarity.  A good candidate for $d_0$ would be
the causal light cone since we should not expect effects impacting the
rate of time evolution to reach beyond the causal light cone.
\begin{figure}[h]
\center
\includegraphics*[width=75mm,height=40mm]{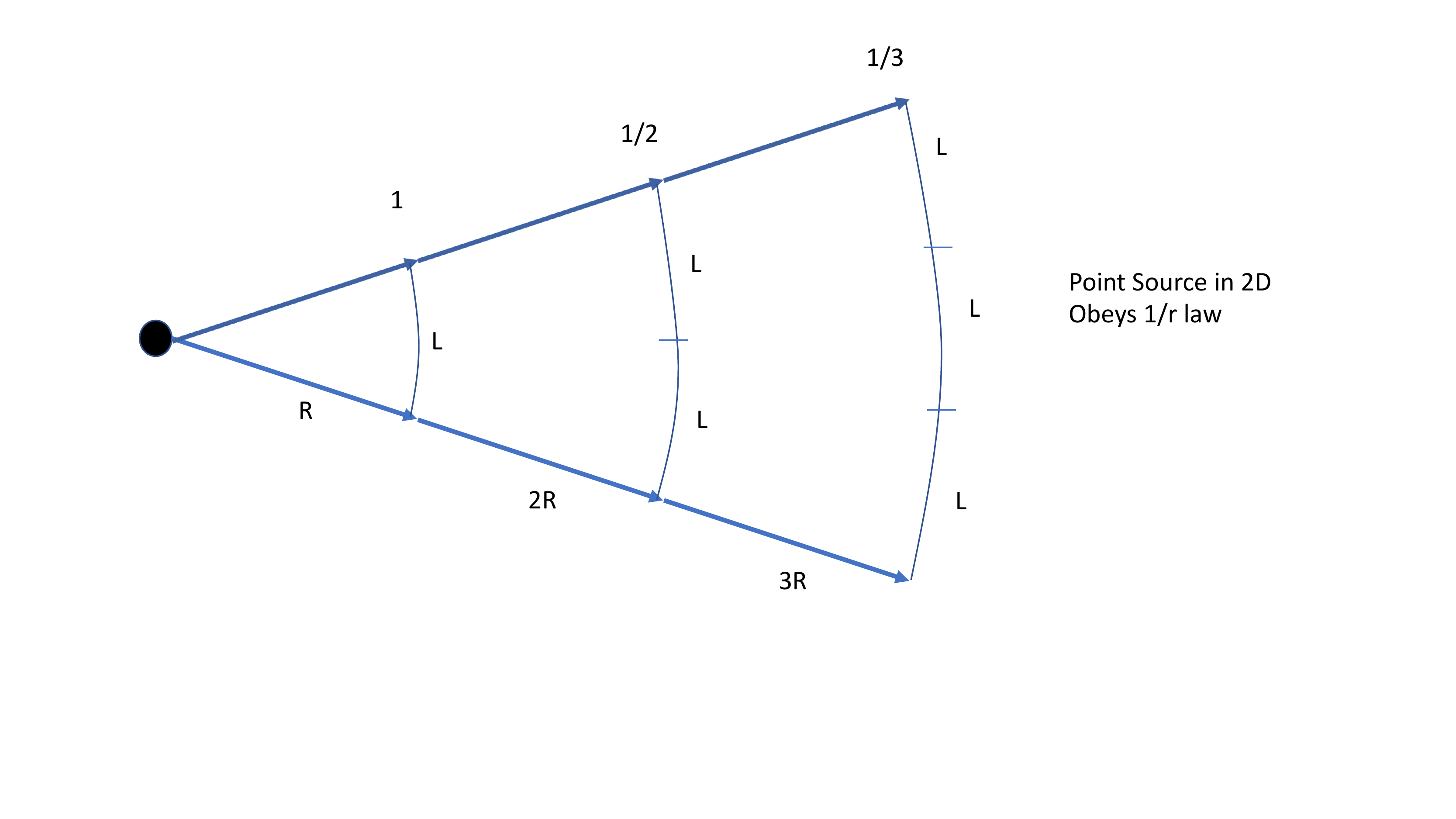}
\caption{The flux of tensor bonds from a point source will fall off as
$\frac{1}{r}$ distance to the source since the flux of entanglement
behaves as a two dimensional object.}
\label{fig4}
\end{figure}
We further conjecture that the potential $aMc^2$  as seen by our test
particle, scales with the  $1/r$ distance to
the center of mass of the object. This is a consequence of Theorem 2
and Postulate 2 and represents our third Theorem:
\begin{ther}
The potential used to calculate the rate of quantum information
propagation over the causal distance $d_0$ is proportional to $1/r$
distance to the center of mass-energy which generates the potential.
\end{ther}

Since by Postulate 2 the  tunneling potential used to calculate
quantum information propagation is proportional to the mass-energy and
by Theorem 2 this same mass-energy is related to the total number of
tensor bonds due to entanglement, then the flux of tensor bonds through an
space should fall as $1/r$ distance to the center of mass-energy. 

The $1/r$ is justified as due to the
flux of entanglement experienced by a test particle at a distance $r$
from a source. A three dimensional radiating point source will observe
a $1/r^2$ flux. However since it has been shown that entanglement
obeys an area law, one expects that entanglement flux will behave as a
two dimensional object (see Fig.~\ref{fig4}).
Thus the entanglement flux will follow instead a $1/r$ law.  In this
case $a M c^2$ becomes $ b \frac{M c^2 }{r }$ with $b$
being an arbitrary proportionality constant having units of length. Now we can
write Eq.~\ref{eq5} as:
\begin{eqnarray}
t_T = \frac{ 1}{i \sqrt{\frac{E_0 }{ 2 d_0^2 m_0 } (1- \frac{b
  M c^2}{r  E_0 } )} } .\label{eq7}
\end{eqnarray}
If we argue that the kinetic zero point energy is equivalent to zero
point mass, $E_0 = m_0 c^2$ we now obtain:
\begin{eqnarray}
t_T = \frac{ 1}{i \sqrt{\frac{c^2}{ 2 d_0^2 } (1- \frac{b
  Mc^2}{r E_0 } )} } .\label{eq7b}
\end{eqnarray}

Equating our minimum time update to Eq.~\ref{eq7} via a constant factor $A_0$ we obtain:
\begin{eqnarray}
\delta t_{min} & = & A_0 t_T \nonumber \\
 & = &  \frac{ 1}{\sqrt{ 1 - \frac{2 G M}{r c^2}}}.
\end{eqnarray}
Here  $ A_0 = \sqrt{\frac{ c^2 }{ 2 d_0^2 }} $ is a time scale factor which has dropped the complex $i$ since Eq.~\ref{eq4} is
  properly done using an absolute value. We find that to make $A_0=1 s^{-1}$
  that $c/\sqrt{2} \times seconds = d_0 = 2.11985\times 10 ^8 m$. So our
  intuition that $d_0$ is related to the causal light cone appears
  justified, only it is scaled by $\sqrt{2}$.
  
Further we have identified $b = \frac{2 G E_0 }{ c^4} $, with $G$ the
  gravitational constant. Using this we can derive a relationship for
  the zero point kinetic energy:
\begin{eqnarray}
E_0 = \frac{b c^4}{2 G}
\end{eqnarray}
If we pick the Planck length for $b= l_p=
\sqrt{\frac{\hbar G}{c^3}}  \approx 1.6\times 10^{-35}$m then we get
$E_0= 9.7\times 10^8 $ J. If we calculate the
energy density by dividing by $l_p^3$ we get $2.36\times 10^{113}
J/m^3$.  This
is almost exactly the high end estimate for vacuum energy density
given using Planck's approach.  In pure constants, the vacuum energy
density becomes:
\begin{eqnarray}
\rho_E = \frac{c^7}{2 G^2 \hbar}.
\end{eqnarray}

Thus using the tunneling time expression, we have recovered the standard time dilation in
a gravitation field:
\begin{eqnarray}
  t_f = \frac{ t_0}{\sqrt{ 1 - \frac{2 G M}{r c^2}}}
\end{eqnarray}
Here $t_f$ is the  is the coordinate time between events for a
fast-ticking observer at an arbitrarily large distance from the
massive object, $G $ is the Gravitational constant, $M$ is the mass of
the object creating the gravitational field and $r$ is the radial
coordinate and $t_0$ the observed time for the observer.

\section{Discussion}
This attempt to derive the mass-energy time dilation equation using the tunneling time
formula from quantum mechanics has the appeal that one can recover a
believable quantum correlation distance proportional to the causal
light cone.   As well as a vacuum energy density consistent with older
and higher estimates is also recovered. This might be significant since a large issue in
reconciling quantum mechanics with General relativity has been
accounting for the large vacuum energy density predicted by quantum
mechanics.  Here the large energy density
follows, as a natural consequence of this derivation.

Starting with the gravitational time dilation
equation one should be able to re-derive Einstein's field
equations. Here the governing idea is that mass-energy slows the
update of quantum states due to the finite time it takes to update
quantum correlations in parallel. It is this differential in time updates
which drives the emergence of the force of gravitation.

\bibliography{ComplexTime}

\providecommand{\noopsort}[1]{}\providecommand{\singleletter}[1]{#1}%
\begin{thebibliography}{20}%
\makeatletter
\providecommand \@ifxundefined [1]{%
 \@ifx{#1\undefined}
}%
\providecommand \@ifnum [1]{%
 \ifnum #1\expandafter \@firstoftwo
 \else \expandafter \@secondoftwo
 \fi
}%
\providecommand \@ifx [1]{%
 \ifx #1\expandafter \@firstoftwo
 \else \expandafter \@secondoftwo
 \fi
}%
\providecommand \natexlab [1]{#1}%
\providecommand \enquote  [1]{``#1''}%
\providecommand \bibnamefont  [1]{#1}%
\providecommand \bibfnamefont [1]{#1}%
\providecommand \citenamefont [1]{#1}%
\providecommand \href@noop [0]{\@secondoftwo}%
\providecommand \href [0]{\begingroup \@sanitize@url \@href}%
\providecommand \@href[1]{\@@startlink{#1}\@@href}%
\providecommand \@@href[1]{\endgroup#1\@@endlink}%
\providecommand \@sanitize@url [0]{\catcode `\\12\catcode `\$12\catcode
  `\&12\catcode `\#12\catcode `\^12\catcode `\_12\catcode `\%12\relax}%
\providecommand \@@startlink[1]{}%
\providecommand \@@endlink[0]{}%
\providecommand \url  [0]{\begingroup\@sanitize@url \@url }%
\providecommand \@url [1]{\endgroup\@href {#1}{\urlprefix }}%
\providecommand \urlprefix  [0]{URL }%
\providecommand \Eprint [0]{\href }%
\providecommand \doibase [0]{http://dx.doi.org/}%
\providecommand \selectlanguage [0]{\@gobble}%
\providecommand \bibinfo  [0]{\@secondoftwo}%
\providecommand \bibfield  [0]{\@secondoftwo}%
\providecommand \translation [1]{[#1]}%
\providecommand \BibitemOpen [0]{}%
\providecommand \bibitemStop [0]{}%
\providecommand \bibitemNoStop [0]{.\EOS\space}%
\providecommand \EOS [0]{\spacefactor3000\relax}%
\providecommand \BibitemShut  [1]{\csname bibitem#1\endcsname}%
\let\auto@bib@innerbib\@empty
\bibitem [{\citenamefont {Camus}\ \emph {et~al.}(2017)\citenamefont {Camus},
  \citenamefont {Yakaboylu}, \citenamefont {Fechner}, \citenamefont {Klaiber},
  \citenamefont {Laux}, \citenamefont {Mi}, \citenamefont {Hatsagortsyan},
  \citenamefont {Pfeifer}, \citenamefont {Keitel},\ and\ \citenamefont
  {Moshammer}}]{PhysRevLett.119.023201}%
  \BibitemOpen
  \bibfield  {author} {\bibinfo {author} {\bibfnamefont {N.}~\bibnamefont
  {Camus}}, \bibinfo {author} {\bibfnamefont {E.}~\bibnamefont {Yakaboylu}},
  \bibinfo {author} {\bibfnamefont {L.}~\bibnamefont {Fechner}}, \bibinfo
  {author} {\bibfnamefont {M.}~\bibnamefont {Klaiber}}, \bibinfo {author}
  {\bibfnamefont {M.}~\bibnamefont {Laux}}, \bibinfo {author} {\bibfnamefont
  {Y.}~\bibnamefont {Mi}}, \bibinfo {author} {\bibfnamefont {K.~Z.}\
  \bibnamefont {Hatsagortsyan}}, \bibinfo {author} {\bibfnamefont
  {T.}~\bibnamefont {Pfeifer}}, \bibinfo {author} {\bibfnamefont {C.~H.}\
  \bibnamefont {Keitel}}, \ and\ \bibinfo {author} {\bibfnamefont
  {R.}~\bibnamefont {Moshammer}},\ }\href {\doibase
  10.1103/PhysRevLett.119.023201} {\bibfield  {journal} {\bibinfo  {journal}
  {Phys. Rev. Lett.}\ }\textbf {\bibinfo {volume} {119}},\ \bibinfo {pages}
  {023201} (\bibinfo {year} {2017})}\BibitemShut {NoStop}%
\bibitem [{\citenamefont {Van~Raamsdonk}(2010)}]{VanRaamsdonk:2010pw}%
  \BibitemOpen
  \bibfield  {author} {\bibinfo {author} {\bibfnamefont {M.}~\bibnamefont
  {Van~Raamsdonk}},\ }\href {\doibase 10.1007/s10714-010-1034-0,
  10.1142/S0218271810018529} {\bibfield  {journal} {\bibinfo  {journal} {Gen.
  Rel. Grav.}\ }\textbf {\bibinfo {volume} {42}},\ \bibinfo {pages} {2323}
  (\bibinfo {year} {2010})},\ \bibinfo {note} {[Int. J. Mod.
  Phys.D19,2429(2010)]},\ \Eprint {http://arxiv.org/abs/1005.3035}
  {arXiv:1005.3035 [hep-th]} \BibitemShut {NoStop}%
\bibitem [{\citenamefont {{Ron Cowen}}(2015)}]{nature}%
  \BibitemOpen
  \bibfield  {author} {\bibinfo {author} {\bibnamefont {{Ron Cowen}}},\
  }\href@noop {} {\enquote {\bibinfo {title} {{The quantum source of
  space-time}},}\ }\bibinfo {howpublished} {{NATURE | NEWS FEATURE}} (\bibinfo
  {year} {2015}),\ \bibinfo {note}
  {\url{https://www.nature.com/news/the-quantum-source-of-space-time-1.18797}}\BibitemShut
  {NoStop}%
\bibitem [{\citenamefont {Maldacena}\ and\ \citenamefont
  {Susskind}(2013)}]{Maldacena:2013xja}%
  \BibitemOpen
  \bibfield  {author} {\bibinfo {author} {\bibfnamefont {J.}~\bibnamefont
  {Maldacena}}\ and\ \bibinfo {author} {\bibfnamefont {L.}~\bibnamefont
  {Susskind}},\ }\href {\doibase 10.1002/prop.201300020} {\bibfield  {journal}
  {\bibinfo  {journal} {Fortsch. Phys.}\ }\textbf {\bibinfo {volume} {61}},\
  \bibinfo {pages} {781} (\bibinfo {year} {2013})},\ \Eprint
  {http://arxiv.org/abs/1306.0533} {arXiv:1306.0533 [hep-th]} \BibitemShut
  {NoStop}%
\bibitem [{\citenamefont {Jacobson}(1995)}]{PhysRevLett.75.1260}%
  \BibitemOpen
  \bibfield  {author} {\bibinfo {author} {\bibfnamefont {T.}~\bibnamefont
  {Jacobson}},\ }\href {\doibase 10.1103/PhysRevLett.75.1260} {\bibfield
  {journal} {\bibinfo  {journal} {Phys. Rev. Lett.}\ }\textbf {\bibinfo
  {volume} {75}},\ \bibinfo {pages} {1260} (\bibinfo {year}
  {1995})}\BibitemShut {NoStop}%
\bibitem [{\citenamefont {Lashkari}\ \emph {et~al.}(2014)\citenamefont
  {Lashkari}, \citenamefont {McDermott},\ and\ \citenamefont
  {Van~Raamsdonk}}]{Lashkari:2013koa}%
  \BibitemOpen
  \bibfield  {author} {\bibinfo {author} {\bibfnamefont {N.}~\bibnamefont
  {Lashkari}}, \bibinfo {author} {\bibfnamefont {M.~B.}\ \bibnamefont
  {McDermott}}, \ and\ \bibinfo {author} {\bibfnamefont {M.}~\bibnamefont
  {Van~Raamsdonk}},\ }\href {\doibase 10.1007/JHEP04(2014)195} {\bibfield
  {journal} {\bibinfo  {journal} {JHEP}\ }\textbf {\bibinfo {volume} {04}},\
  \bibinfo {pages} {195} (\bibinfo {year} {2014})},\ \Eprint
  {http://arxiv.org/abs/1308.3716} {arXiv:1308.3716 [hep-th]} \BibitemShut
  {NoStop}%
\bibitem [{\citenamefont {Lin}\ \emph {et~al.}(2015)\citenamefont {Lin},
  \citenamefont {Marcolli}, \citenamefont {Ooguri},\ and\ \citenamefont
  {Stoica}}]{PhysRevLett.114.221601}%
  \BibitemOpen
  \bibfield  {author} {\bibinfo {author} {\bibfnamefont {J.}~\bibnamefont
  {Lin}}, \bibinfo {author} {\bibfnamefont {M.}~\bibnamefont {Marcolli}},
  \bibinfo {author} {\bibfnamefont {H.}~\bibnamefont {Ooguri}}, \ and\ \bibinfo
  {author} {\bibfnamefont {B.}~\bibnamefont {Stoica}},\ }\href {\doibase
  10.1103/PhysRevLett.114.221601} {\bibfield  {journal} {\bibinfo  {journal}
  {Phys. Rev. Lett.}\ }\textbf {\bibinfo {volume} {114}},\ \bibinfo {pages}
  {221601} (\bibinfo {year} {2015})}\BibitemShut {NoStop}%
\bibitem [{\citenamefont {Verlinde}(2017)}]{Verlinde:2016toy}%
  \BibitemOpen
  \bibfield  {author} {\bibinfo {author} {\bibfnamefont {E.~P.}\ \bibnamefont
  {Verlinde}},\ }\href {\doibase 10.21468/SciPostPhys.2.3.016} {\bibfield
  {journal} {\bibinfo  {journal} {SciPost Phys.}\ }\textbf {\bibinfo {volume}
  {2}},\ \bibinfo {pages} {016} (\bibinfo {year} {2017})},\ \Eprint
  {http://arxiv.org/abs/1611.02269} {arXiv:1611.02269 [hep-th]} \BibitemShut
  {NoStop}%
\bibitem [{\citenamefont {Bekenstein}(1973)}]{Bekenstein:1973ur}%
  \BibitemOpen
  \bibfield  {author} {\bibinfo {author} {\bibfnamefont {J.~D.}\ \bibnamefont
  {Bekenstein}},\ }\href {\doibase 10.1103/PhysRevD.7.2333} {\bibfield
  {journal} {\bibinfo  {journal} {Phys. Rev.}\ }\textbf {\bibinfo {volume}
  {D7}},\ \bibinfo {pages} {2333} (\bibinfo {year} {1973})}\BibitemShut
  {NoStop}%
\bibitem [{\citenamefont {Margolus}\ and\ \citenamefont
  {Levitin}(1998)}]{Margolus:1997ih}%
  \BibitemOpen
  \bibfield  {author} {\bibinfo {author} {\bibfnamefont {N.}~\bibnamefont
  {Margolus}}\ and\ \bibinfo {author} {\bibfnamefont {L.~B.}\ \bibnamefont
  {Levitin}},\ }\bibfield  {booktitle} {\emph {\bibinfo {booktitle} {{4th
  Workshop on Physics and Computation (PhysComp 96) Boston, Massachusetts,
  November 22-24, 1996}}},\ }\href {\doibase 10.1016/S0167-2789(98)00054-2}
  {\bibfield  {journal} {\bibinfo  {journal} {Physica}\ }\textbf {\bibinfo
  {volume} {D120}},\ \bibinfo {pages} {188} (\bibinfo {year} {1998})},\ \Eprint
  {http://arxiv.org/abs/quant-ph/9710043} {arXiv:quant-ph/9710043 [quant-ph]}
  \BibitemShut {NoStop}%
\bibitem [{\citenamefont {Brown}\ \emph {et~al.}(2016)\citenamefont {Brown},
  \citenamefont {Roberts}, \citenamefont {Susskind}, \citenamefont {Swingle},\
  and\ \citenamefont {Zhao}}]{Brown:2015lvg}%
  \BibitemOpen
  \bibfield  {author} {\bibinfo {author} {\bibfnamefont {A.~R.}\ \bibnamefont
  {Brown}}, \bibinfo {author} {\bibfnamefont {D.~A.}\ \bibnamefont {Roberts}},
  \bibinfo {author} {\bibfnamefont {L.}~\bibnamefont {Susskind}}, \bibinfo
  {author} {\bibfnamefont {B.}~\bibnamefont {Swingle}}, \ and\ \bibinfo
  {author} {\bibfnamefont {Y.}~\bibnamefont {Zhao}},\ }\href {\doibase
  10.1103/PhysRevD.93.086006} {\bibfield  {journal} {\bibinfo  {journal} {Phys.
  Rev.}\ }\textbf {\bibinfo {volume} {D93}},\ \bibinfo {pages} {086006}
  (\bibinfo {year} {2016})},\ \Eprint {http://arxiv.org/abs/1512.04993}
  {arXiv:1512.04993 [hep-th]} \BibitemShut {NoStop}%
\bibitem [{\citenamefont {Moreva}\ \emph {et~al.}(2014)\citenamefont {Moreva},
  \citenamefont {Brida}, \citenamefont {Gramegna}, \citenamefont {Giovannetti},
  \citenamefont {Maccone},\ and\ \citenamefont {Genovese}}]{Moreva:2013ska}%
  \BibitemOpen
  \bibfield  {author} {\bibinfo {author} {\bibfnamefont {E.}~\bibnamefont
  {Moreva}}, \bibinfo {author} {\bibfnamefont {G.}~\bibnamefont {Brida}},
  \bibinfo {author} {\bibfnamefont {M.}~\bibnamefont {Gramegna}}, \bibinfo
  {author} {\bibfnamefont {V.}~\bibnamefont {Giovannetti}}, \bibinfo {author}
  {\bibfnamefont {L.}~\bibnamefont {Maccone}}, \ and\ \bibinfo {author}
  {\bibfnamefont {M.}~\bibnamefont {Genovese}},\ }\href {\doibase
  10.1103/PhysRevA.89.052122} {\bibfield  {journal} {\bibinfo  {journal} {Phys.
  Rev.}\ }\textbf {\bibinfo {volume} {A89}},\ \bibinfo {pages} {052122}
  (\bibinfo {year} {2014})},\ \Eprint {http://arxiv.org/abs/1310.4691}
  {arXiv:1310.4691 [quant-ph]} \BibitemShut {NoStop}%
\bibitem [{\citenamefont {DeWitt}(1967)}]{PhysRev.160.1113}%
  \BibitemOpen
  \bibfield  {author} {\bibinfo {author} {\bibfnamefont {B.~S.}\ \bibnamefont
  {DeWitt}},\ }\href {\doibase 10.1103/PhysRev.160.1113} {\bibfield  {journal}
  {\bibinfo  {journal} {Phys. Rev.}\ }\textbf {\bibinfo {volume} {160}},\
  \bibinfo {pages} {1113} (\bibinfo {year} {1967})}\BibitemShut {NoStop}%
\bibitem [{\citenamefont {Kitada}(1994)}]{Kitada:1994dq}%
  \BibitemOpen
  \bibfield  {author} {\bibinfo {author} {\bibfnamefont {H.}~\bibnamefont
  {Kitada}},\ }\href {\doibase 10.1007/BF02727290} {\bibfield  {journal}
  {\bibinfo  {journal} {Nuovo Cim.}\ }\textbf {\bibinfo {volume} {B109}},\
  \bibinfo {pages} {281} (\bibinfo {year} {1994})},\ \Eprint
  {http://arxiv.org/abs/astro-ph/9309051} {arXiv:astro-ph/9309051 [astro-ph]}
  \BibitemShut {NoStop}%
\bibitem [{\citenamefont {Dugić}\ and\ \citenamefont
  {Ćirković}(2002)}]{DUGIC2002291}%
  \BibitemOpen
  \bibfield  {author} {\bibinfo {author} {\bibfnamefont {M.}~\bibnamefont
  {Dugić}}\ and\ \bibinfo {author} {\bibfnamefont {M.~M.}\ \bibnamefont
  {Ćirković}},\ }\href {\doibase
  https://doi.org/10.1016/S0375-9601(02)01198-2} {\bibfield  {journal}
  {\bibinfo  {journal} {Physics Letters A}\ }\textbf {\bibinfo {volume}
  {302}},\ \bibinfo {pages} {291 } (\bibinfo {year} {2002})}\BibitemShut
  {NoStop}%
\bibitem [{\citenamefont {Leichenauer}\ \emph {et~al.}(2018)\citenamefont
  {Leichenauer}, \citenamefont {Levine},\ and\ \citenamefont
  {Shahbazi-Moghaddam}}]{Leichenauer:2018obf}%
  \BibitemOpen
  \bibfield  {author} {\bibinfo {author} {\bibfnamefont {S.}~\bibnamefont
  {Leichenauer}}, \bibinfo {author} {\bibfnamefont {A.}~\bibnamefont {Levine}},
  \ and\ \bibinfo {author} {\bibfnamefont {A.}~\bibnamefont
  {Shahbazi-Moghaddam}},\ }\href@noop {} {\  (\bibinfo {year} {2018})},\
  \Eprint {http://arxiv.org/abs/1802.02584} {arXiv:1802.02584 [hep-th]}
  \BibitemShut {NoStop}%
\bibitem [{\citenamefont {Linden}\ \emph {et~al.}(2009)\citenamefont {Linden},
  \citenamefont {Popescu}, \citenamefont {Short},\ and\ \citenamefont
  {Winter}}]{Linden:2009gfy}%
  \BibitemOpen
  \bibfield  {author} {\bibinfo {author} {\bibfnamefont {N.}~\bibnamefont
  {Linden}}, \bibinfo {author} {\bibfnamefont {S.}~\bibnamefont {Popescu}},
  \bibinfo {author} {\bibfnamefont {A.~J.}\ \bibnamefont {Short}}, \ and\
  \bibinfo {author} {\bibfnamefont {A.}~\bibnamefont {Winter}},\ }\href
  {\doibase 10.1103/PhysRevE.79.061103} {\bibfield  {journal} {\bibinfo
  {journal} {Phys. Rev.}\ }\textbf {\bibinfo {volume} {E79}},\ \bibinfo {pages}
  {061103} (\bibinfo {year} {2009})},\ \Eprint {http://arxiv.org/abs/0812.2385}
  {arXiv:0812.2385 [quant-ph]} \BibitemShut {NoStop}%
\bibitem [{\citenamefont {B\"uttiker}(1983)}]{PhysRevB.27.6178}%
  \BibitemOpen
  \bibfield  {author} {\bibinfo {author} {\bibfnamefont {M.}~\bibnamefont
  {B\"uttiker}},\ }\href {\doibase 10.1103/PhysRevB.27.6178} {\bibfield
  {journal} {\bibinfo  {journal} {Phys. Rev. B}\ }\textbf {\bibinfo {volume}
  {27}},\ \bibinfo {pages} {6178} (\bibinfo {year} {1983})}\BibitemShut
  {NoStop}%
\bibitem [{\citenamefont {Landsman}\ \emph {et~al.}(2014)\citenamefont
  {Landsman}, \citenamefont {Weger}, \citenamefont {Maurer}, \citenamefont
  {Boge}, \citenamefont {Ludwig}, \citenamefont {Heuser}, \citenamefont
  {Cirelli}, \citenamefont {Gallmann},\ and\ \citenamefont
  {Keller}}]{Landsman:14}%
  \BibitemOpen
  \bibfield  {author} {\bibinfo {author} {\bibfnamefont {A.~S.}\ \bibnamefont
  {Landsman}}, \bibinfo {author} {\bibfnamefont {M.}~\bibnamefont {Weger}},
  \bibinfo {author} {\bibfnamefont {J.}~\bibnamefont {Maurer}}, \bibinfo
  {author} {\bibfnamefont {R.}~\bibnamefont {Boge}}, \bibinfo {author}
  {\bibfnamefont {A.}~\bibnamefont {Ludwig}}, \bibinfo {author} {\bibfnamefont
  {S.}~\bibnamefont {Heuser}}, \bibinfo {author} {\bibfnamefont
  {C.}~\bibnamefont {Cirelli}}, \bibinfo {author} {\bibfnamefont
  {L.}~\bibnamefont {Gallmann}}, \ and\ \bibinfo {author} {\bibfnamefont
  {U.}~\bibnamefont {Keller}},\ }\href {\doibase 10.1364/OPTICA.1.000343}
  {\bibfield  {journal} {\bibinfo  {journal} {Optica}\ }\textbf {\bibinfo
  {volume} {1}},\ \bibinfo {pages} {343} (\bibinfo {year} {2014})}\BibitemShut
  {NoStop}%
\bibitem [{\citenamefont {Pollak}\ and\ \citenamefont
  {Miller}(1984)}]{PhysRevLett.53.115}%
  \BibitemOpen
  \bibfield  {author} {\bibinfo {author} {\bibfnamefont {E.}~\bibnamefont
  {Pollak}}\ and\ \bibinfo {author} {\bibfnamefont {W.~H.}\ \bibnamefont
  {Miller}},\ }\href {\doibase 10.1103/PhysRevLett.53.115} {\bibfield
  {journal} {\bibinfo  {journal} {Phys. Rev. Lett.}\ }\textbf {\bibinfo
  {volume} {53}},\ \bibinfo {pages} {115} (\bibinfo {year} {1984})}\BibitemShut
  {NoStop}%
\end{thebibliography}%

\end{document}